# CFD-Driven Optimization of Dual-Throat Micro-Nozzle for Enhanced Thrust in Cold Gas Propulsion for Microsatellites

AmirHossein Niksirat [a], *

*a. Ph.D. Researcher, Tehran, Iran*

**Abstract:** Micro spacecraft have garnered increasing favor for their adaptability in space missions. Cold gas propellant-based (CGP) micro propulsion systems assume a pivotal role in enabling maneuvering, monitoring, and precise attitude control in micro/nanosatellites, offering attributes of simplicity, non-toxicity, and minimal leakage compared to alternative propulsion methods. The convergent-divergent micro-nozzle, an integral component of micro thrusters, is instrumental in thrust generation. This research conducts a sensitivity analysis of the micro-nozzle to discern influential geometric variables, employing response surface optimization techniques through ANSYS FLUENT. Results indicate a noteworthy thrust enhancement, reaching 113.1 mN in the optimized design. Additionally, a novel curved throat concept is introduced, yielding a 25 percent thrust increase and culminating in a final design achieving 141 mN. Further exploration involves a dual-throat micro-nozzle, integrating response surface optimization and a curved throat design, resulting in an impressive thrust level of 261 mN, surpassing prior designs without added complexity. This study advances micro spacecraft propulsion by addressing challenges and introducing innovative approaches, emphasizing the pioneering nature of this investigation within the realm of micro propulsion systems.

**Keywords:** Micro-thruster; Cold gas propulsion; Dual-throat micro-nozzle; CFD; Optimization; Design of experiments

*Corresponding author. Tel.: (+98) 9125139762. E-mail address: amirhossien.niksirat@gmail.com.

## Nomenclature and abbreviations

CGP: cold gas propulsion

### Nomenclature

| | |
|---|---|
| $w_t$ | throat width (unit: $\mu m$) |
| $w_e$ | exit width (unit: $\mu m$) |
| $w_i$ | inlet width (unit: $\mu m$) |
| $F$ | thrust (unit: mN) |
| $\in$ | expansion ratio |
| $\alpha$ | semi convergent angle (unit: degree) |
| $\alpha_2$ | second-semi convergent angle (unit: degree) |
| $\beta$ | semi divergent angle (unit: degree) |
| $V_e$ | exit velocity (m/s) |
| $p_c$ | chamber pressure (unit: bar) |
| $p_e$ | exit pressure (unit: bar) |
| $P_a$ | ambient pressure (unit: bar) |
| $\dot{m}$ | mass flow rate (unit: $mgs^{-1}$) |
| $L_{conv}$ | convergence length (unit: $\mu m$) |
| $L_{conv_1}$ | second convergence length (unit: $\mu m$) |
| $L_d$ | divergence length (unit: $\mu m$) |
| $I_{sp}$ | specific impulse (unit: s) |
| $C_f$ | thrust correction factor |
| $C^*$ | characteristic velocity |

### Subscripts

| | |
|---|---|
| $i$ | inlet |
| $e$ | exit |
| $c$ | chamber |
| $t$ | throat |
| $conv$ | convergence |
| $div$ | divergence |
| $sp$ | specific impulse |
| $f$ | factor |

## 1. Introduction

According to various economic reports, the global market for small satellites is growing rapidly in the next decade. Demands for small satellites have increased in various industries [1, 2]. Many companies and organizations as well as start-ups ramp up their efforts to produce many small satellites and clamp down on the excessive use of large satellites [3, 4]. The micro propulsion system is an active field in research on micro/nanosatellites. Modern artificial satellites are used to collect information on a large scale around the earth. In this regard, a well-known class of small satellites has become popular due to their cost-effectiveness [5, 6]. The use of microelectromechanical systems (MEMS) in micro-propulsion systems and microfabrication technologies in micro dimensions have created potential in aerospace technology. NASA's comparison between the benefits of micro thrusters and reaction wheels on future observatory-class missions has demonstrated tight-pointing stability [7]. Typically, the propulsion subsystem accounts for 10% of the total satellite mass, and this amount is directly related to the cargo mass, power, and volume required, and reducing them has a crucial effect on the cost of the mission [8]. Any propulsion system that can be used for a microsatellite with a mass of less than 100 kg is called a micro propulsion system. The term micro thruster is used to denote a propulsion system with micro to a milli-Newton and has no relation with the size of the satellite. Thus, a micro propulsion system is suitable not only for micro/nanosatellites but also for larger satellites that require more precise equipment in terms of accuracy and stability. Recent surveys report the impeccable growth of small satellites during the past decade. Figure 1 illustrates the launch growth of nanosatellites over the years and predicts it up to 2027. The integration of small satellites is crucial for shaping the future of space applications, spanning diverse areas such as earth observation/remote sensing, technology, scientific endeavors, communications, and novel applications. In the meantime, the importance of the presence of micro/nanosatellites in communications is evident [9-11]. The thrust that is gained in micro thruster is from the micro to milli-Newton range to attain the requirements of maneuvering, attitude, station keeping, drag compensation, and orbit adjustment [12-14].

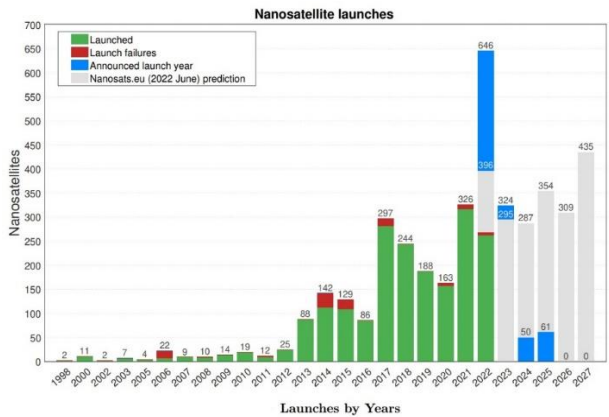

**Figure 1** The trajectory of nanosatellite deployment over successive years, with prognostications extending through 2027 [9].

The inception of micro propulsion using MEMS technology traces back to Mitterauer in 1991, introducing Micro Field Emission Electric Propulsion (FEEP) based on Field Emitter Array (FEA) technology. The initial goal was to design a compact system with micro-meter-sized components, but no laboratory models were developed at that time. Subsequently, in 1994, the Janson Aerospace Center explored MEMS-based propulsion, including resistive jets and ion propulsion. Concurrently, ACR, supported by the European Space Agency, initiated studies on MEMS-based cold gas propellant-based (CGP) micro propulsion. Simultaneously, the French National Research Center investigated an array of micro-machined solid engines. These early endeavors laid the foundation for ongoing micro thruster projects across various institutions. To contribute to this evolving field, our focus is on reviewing recent papers, identifying their limitations, and proposing enhancements [15].

Micro propulsion in space missions falls into two main categories: chemical and electrical. Chemical micro propulsion operates on the common principle of propellant release to achieve reaction force, with fuel options including solid, liquid, or gas. The simplest form is the cold gas propellant-based (CGP) thruster, utilizing gas expulsion through a micro-nozzle without undergoing chemical reactions. In electric micro propulsion, positively charged ions generate the driving force. These systems are a product of MEMS technology innovation in space applications.

Given their intrinsic advantages—namely simplicity, non-toxicity, absence of combustion, and minimal leakage—CGP systems have emerged as pivotal elements in micro/nanosatellites. The quintessential micro propulsion system seeks to minimize mass and complexity while optimizing thrust precision. Nonetheless, cold gas micro thrusters grapple with challenges pertaining to weight and high-pressure considerations [13,16]. The schematic of the cold gas micro thruster is shown in Figure 2.

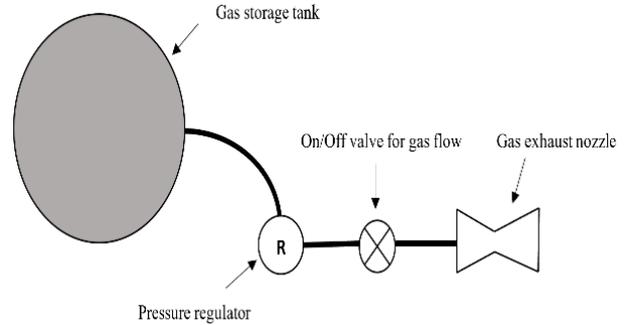

**Figure 2** Comprehensive diagram illustrating the architecture of a cold gas micro thruster.

In terms of hardware intricacy, cold gas micro-thrusters exhibit a comparative simplicity when contrasted with pulsed plasma thrusters, colloidal thrusters, and field emission electric propulsion thrusters. Fundamental constituents include a microvalve, nozzle, and storage system, with the desired thrust attained through the controlled release of compressed gas. Reichbach [17] conducted an investigation to determine the size and design requirements of micro-thrusters, including mass flow, valve dimensions, feed system, and storage tank details. The results indicate that the response time of micro-thrusters is longer than the required response time to achieve the desired thrust value. Consequently, it can be concluded that micro-thrusters are not suitable for attitude control, but they are adequate for orbit transfer and maneuvering purposes. During the development of micro-thrusters, different groups such as NASA, VACCO Industries, CheMS Technology, UTIAS, Stanford's SSD Labs, and Uppsala University have devised various valve configurations [18-20]. Noci et al. [23] reported the development of a micro propulsion feed module for a CGP (Cold Gas Propulsion) system aimed at achieving precise pointing of a scientific satellite. Furthermore, Noci et al. [23] developed the micro propulsion subsystem to compensate for environmental disturbances in the GAIA spacecraft by controlling the thrust through valve throttling.

Cold gas micro thrusters, frequently employed in the orchestration of satellite attitude, exhibit commendable traits. These encompass heightened reliability, diminished system intricacy without ignition interference, secure operational characteristics, absence of exhaust pollution on external satellite surfaces, and a relatively modest impulse. Another noteworthy aspect of cold gas micro thrusters lies in their applicability to missions necessitating minimal velocity alterations [2]. In Table 1, we provide a summary of suggested applications for the thrust versus specific impulse.

**Table 1** Suggested applications for the Thrust versus Specific impulse [15].

| Thrust | Specific impulse | Suggested application |
|--------|------------------|----------------------|
| High | Low | space debris removal, fast orbital transfer/maneuvers (When spacecraft stability is not an issue) |
| Low | High | precise pointing, slow orbital transfer/maneuvers |
| Low | low | attitude control, small orbit corrections (max. ΔV in the order of a few m/s) |

As previously emphasized, the selection of a micro thruster system for spacecraft or satellites is contingent upon the specific mission and application. Nevertheless, certain criteria hold universal significance and demand meticulous evaluation across all missions. These considerations encompass the critical assessment of system performance, safety and reliability parameters, controllability aspects, size and volume adaptability within the confined space of the satellite or spacecraft, usability considerations, fabrication cost implications, and power consumption constraints for thrust generation. In the evaluation of micro thruster systems, factors such as the thrust time profile, the number of impulses available for mission optimization, the inherent risk of failure or damage, the precision of controllability over the ignition process, response time to commands, and the overall efficiency in terms of ease of use assume pivotal importance [1].

Several types of research have been conducted to optimize and improve the performance of cold gas micro thrusters by different groups in the world. The design of CGP systems employs various propellants including hydrogen, helium, nitrogen, argon, and methane as gaseous propellants and ammonia and carbon dioxide as a liquid propellant, and uses eight different material options, including titanium, titanium alloy, magnesium alloy, beryllium alloy, steel, carbon fiber composites, glass fiber composites, and aluminum alloy were investigated. Three optimization methods were used to scrutinize the optimal chamber pressure, the optimal total mass of the system, the radius of the propellant tank, and the principles of structural design. Nitrogen as the preferred propellant and composite due to its lightweight for the CGP system was proposed [2, 14].

This Cold Gas Propulsion (CGP) concept aims to enhance efficacy by focusing on pivotal components, including the gas storage tank, pressure regulator, micro valve, and micro-nozzle. The investigation prioritizes the micro-nozzle, recognizing its substantial influence on thrust. To optimize thrust generation, a comprehensive simulation and optimization of the micro-nozzle have been executed. Figure 3 illustrates the various components comprising the cold gas generator.

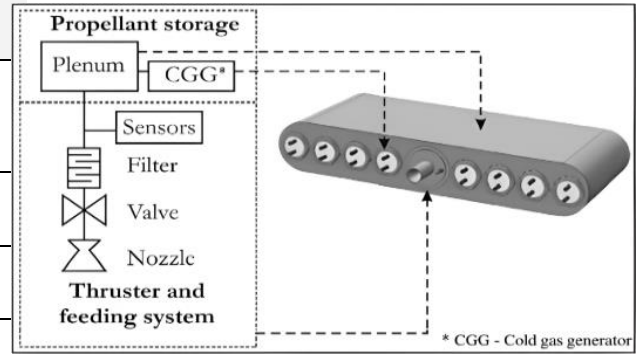

**Figure 3** An illustration of the Cold Gas Propulsion (CGP) thruster components devised for the Delfi-N3xt mission [17].

## 2. Design and numerical model

This work aims to design and optimize the C-D micro-nozzle of the micro propulsion system using compressed air as a propellant. These types of C-D micro-nozzles are similar to the strained tube to form an hourglass shape called the throat. The purpose of this section is to accelerate the passing flow of propellant to supersonic velocity [2]. Several

Variables play important roles in determining the performance of the C-D micro-nozzle. To investigate the impact of these variables on the performance of the C-D micro thruster and enhance the generated thrust, numerous simulations were conducted for sensitivity analysis of geometrical variables. These variables include the inlet width, outlet width, throat width, and convergence and divergence lengths. To model the C-D micro thruster system, a commercial finite volume method (FVM) software, ANSYS computational fluid dynamics, was utilized.

*2.1. Basic Assumptions*

To generate the FVM model of the C-D micro thruster, some assumptions were considered and the simulation was done in steady state, isentropic, and quasi-flow conditions:

**Table 2** Geometrical variables of the micro thruster.

| Inlet width | Throat width | Exit diameter | Semi convergent | Semi divergent | Convergence length | Divergence length |
|---|---|---|---|---|---|---|
| $w_i(\mu m)$ | $w_t(\mu m)$ | $w_e(\mu m)$ | $\alpha(°)$ | $\beta(°)$ | $L_{conv}(\mu m)$ | $L_d(\mu m)$ |
| 1500 | 30 | 450 | 28 | 28 | 2764.67 | 789.91 |

1) The passing flow through the C-D micro-nozzle is adiabatic and the propellant is an ideal gas and compressible.
2) No friction and no boundary layer are considered.
3) The leaving flow is only axial.
4) Gas physical properties (pressure, velocity, temperature, and density) are uniform for each cross-section.
5) No chemical reaction [3].

In the context of scaled-down C-D micro-nozzles, the increased surface-to-volume ratio is attributed to heightened friction. In these micro-nozzles, the boundary layer is observed to extend throughout the entire structure, prompting the application of inviscid flow theory for precise performance predictions. Recently, cold gas micro thrusters have been extensively developed to generate small thrust levels to utilize in microsatellites for attitude control and also position control due to their unique advantages. In addition, simulation techniques due to the high Knudsen number flow within the C-D micro-nozzle that speedily expands to rarefied gas are quite complicated [4].

The geometrical variables used for calculating thrust are provided in Table 2. It is important to note that these variables have varying effects on the resulting thrust. Appendix contains the equations used to calculate the theoretical values of the Mach number, temperature, pressure, and jet velocity at the exit of the C-D micro thruster. Additionally, it includes information on the choked mass flow rate, thrust produced, characteristic velocity, thrust coefficient, and specific impulse. For a visual representation of the main variables, please refer to Figure 4 depicting the CGP's key parameters.

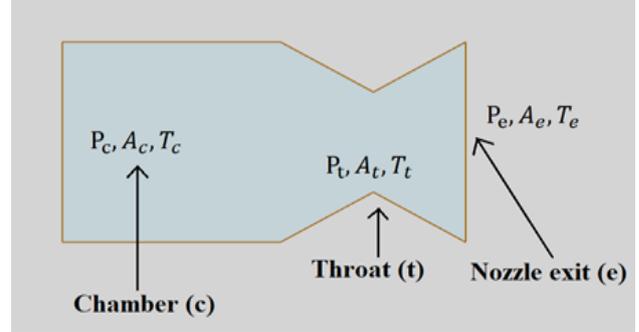

**Figure 4** The fundamental components in the cold gas propulsion system of the C-D micro thruster are the Chamber, Throat, and Exit, constituting the most crucial variables associated with pressure, area, and temperature.

### 2.2. Numerical model

The specified model, utilizing the mentioned dimensions, was developed. Simulations were performed for inlet pressure and environmental conditions using the commercial software ANSYS Fluent [21]. Validation for mesh dependency was conducted to ensure reasonable accuracy and was subsequently approved. The most common and highly effective k-ε turbulence model is used for simulating mean flow characteristics, shock, and discontinuity throw the C-D micro-nozzle. C-D micro-nozzle aero-thermodynamic properties including velocity contour, Mach number contour, pressure contour, and the corresponding velocity vector contour in the C-D micro-nozzle [19] are generated. C-D micro-nozzle geometry drawn in the ANSYS Workbench is also shown in Figure 5. Simulation has been performed for a specific case and solver setup details are shown in Table 3.

**Table 3** Fluent solver setup

| Problem setup | |
|---|---|
| General | Solver type: density – based 2D<br>Space: axisymmetric<br>Steady state |
| Models | Energy equation: on<br>Viscous model: k-ε model with standard wall treatment and compressibility effect |
| Materials | Ideal gas-air |
| Boundary Conditions | Inlet pressure: 2-bar ΔP<br>Set supersonic initial gauge pressure<br>Turbulent intensity: 5%<br>Turbulent viscosity: 10<br>Thermal: 300 k<br>No-slip wall<br>Operating pressure: 0<br>Outlet pressure: sea level – 1 bar |
| Solution setup | |
| Solution controls | Courant number: 0.5 |
| Solution initialization | Compute from: Inlet<br>Convergence criteria: 1e−6 |
| Run Calculation | Check case, enter number of iterations: 1000 |
| Results | |
| Graphics and animation | CFD-Post |

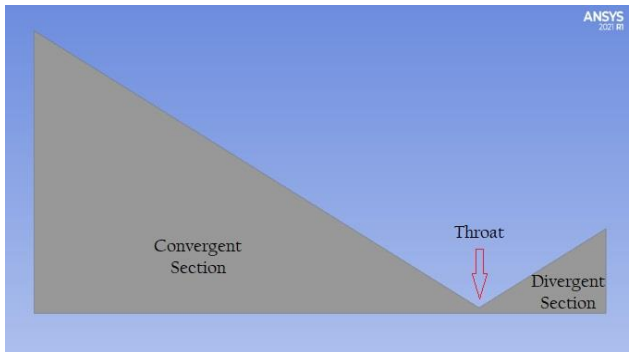

**Figure 5** C-D micro-nozzle geometry drawn in the ANSYS Workbench.

## 2.3. Boundary condition

The boundary condition for sea level (1 bar absolute) conditions was specified 2-bar ΔP (bar) between chamber pressure and ambient pressure. Additionally, employing a convergence criterion with a residual of 1e-6, we successfully obtained values for velocity, pressure, Mach number, and velocity vector.

## 2.4. Numerical results for sea level condition

To verify the simulations' results, the experimental results published in Ref [25], were used as a reference, and the simulation results conducted in this paper were compared. The results showed a good and reasonable agreement between the experimental results and simulation results which approve and verify the simulation details and process. In this family of simulations, based on specific conditions that were defined, the C-D micro-nozzle was simulated in order to analyze the variation of velocity, Mach number, pressure, and velocity vector and compare it with the experimental results.

The findings depicted in Figure 6 demonstrate a sudden increase in velocity and Mach number upon passing the throat, attributed to the reduction in area. The reverse flow hampers the exit velocity, resulting in diminished thrust at the exit section. Figure 7.a illustrates a decrease in pressure caused by the narrowing of the throat area and an increase in throat velocity. Negligible pressure changes are observed at the exit section. Figure 7.b depicts variations in the velocity vector. The adverse impact of reverse flow on C-D micro-nozzle performance is evident, leading to a reduction in thrust generation. To improve the micro thruster's performance, it is essential to minimize or eliminate reverse flow. This paper will discuss one of the most effective approaches, which involves optimizing the C-D micro-nozzle geometry in the subsequent section.

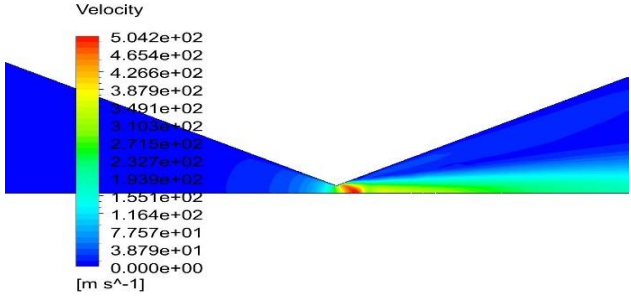

(a) Velocity variation for a 2-bar pressure

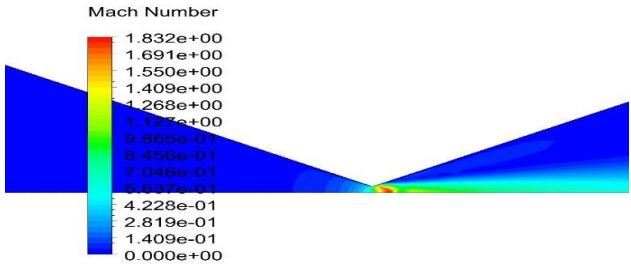

(b) Mach number variation for a 2-bar pressure

**Figure 6** velocity variation and Mach number variation for a 2-bar pressure difference under sea level conditions.

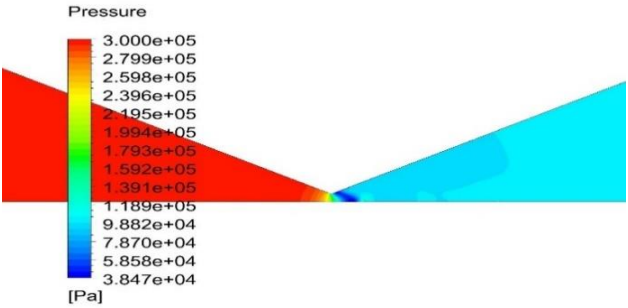

(a) Variation in pressure for a 2-bar pressure

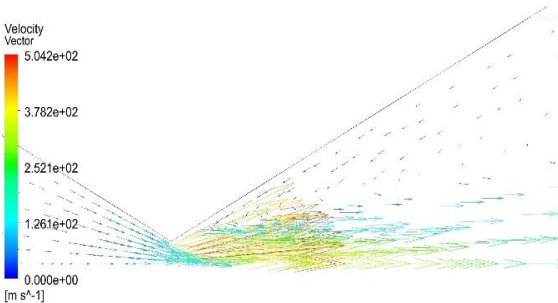

(b) variation in velocity vector for a 2-bar pressure

**Figure 7** Variation in pressure and variation in velocity vector for a 2-bar pressure difference under sea level conditions.

## 3. Response surface optimization

Gaining a comprehensive understanding of the impact of each variable on micro thruster performance offers significant advantages, as it enables identification of the most influential variables for optimizing thrust generation. This section presents a two-step approach: sensitivity analysis, based on a large number of simulations, followed by the application of the response surface optimization method to enhance the micro thruster's geometry and maximize thrust output. The response surface optimization method comprises three stages:

1) Design of Experiments (DOE): A carefully planned set of experiments is conducted.

2) Response Surface: Based on the results from the DOE, a mathematical model or response surface is created to capture the relationship between the variables and the desired performance metrics.

3) Optimization: The response surface is utilized to determine the optimal values of the variables that will yield the highest thrust output.

### 3.1. Design of experiments

Here, we determine the minimum and maximum values of each input variable by considering the forced design constraints and incorporating the application of micro thrusters (Table 4). Subsequently, we employ the Central Composite Design method to establish the number of states or points. At each of these states or points, we calculate the thrust.

The flow passing through the C-D micro-nozzle is adiabatic, and the propellant is assumed to be an ideal gas with compressible properties.

**Table 4** Variation ranges of C-D micro-nozzle variables.

| Variable | Range (min to max) |
| --- | --- |
| Inlet width $W_i$ ($\mu m$) | 300 to 1500 |
| Exit width $W_e$ ($\mu m$) | 40 to 1500 |
| Throat width $W_t$ ($\mu m$) | 6 to 500 |
| Convergence length $L_{conv}$ ($\mu m$) | 1300 t0 4200 |
| Divergence length $L_d$ ($\mu m$) | 300 to 1200 |

**Table 5** Values of response point variables.

| Variable | Inlet width | Exit width | Throat width | Convergence length | Divergence length | Thrust |
|---|---|---|---|---|---|---|
| Response point | $W_i$ ($\mu m$) | $w_e$ ($\mu m$) | $W_t$ ($\mu m$) | $L_{conv}$ ($\mu m$) | $L_d$ ($\mu m$) | $F$ (mN) |
| | 900 | 770 | 253 | 2750 | 750 | 48 |

*3.2. Response surface*

In this stage, the data obtained from the previous step is updated, and the Genetic Aggregation model is used to determine a point on the response surface, as presented in Table 5. The sensitivity of different variables (inlet width, exit width, throat width, convergence length, and divergence length) of the C-D micro-nozzle is depicted in the pie chart shown in Figure 8. It is evident that throat width and exit width are the most influential variables among the others, while the inlet width has a lesser effect on the generated thrust. Figure 9 and Figure 10 demonstrate a linear relationship between thrust and inlet width and throat width, unlike the exit width. The variation of the exit width generally exhibits an upward trend, and its optimal point value needs to be determined (Figure 11). Similarly, the thrust variation as a function of the divergence and convergence lengths of the C-D micro-nozzle follows a similar pattern, as observed in Figure 12 and Figure 13.

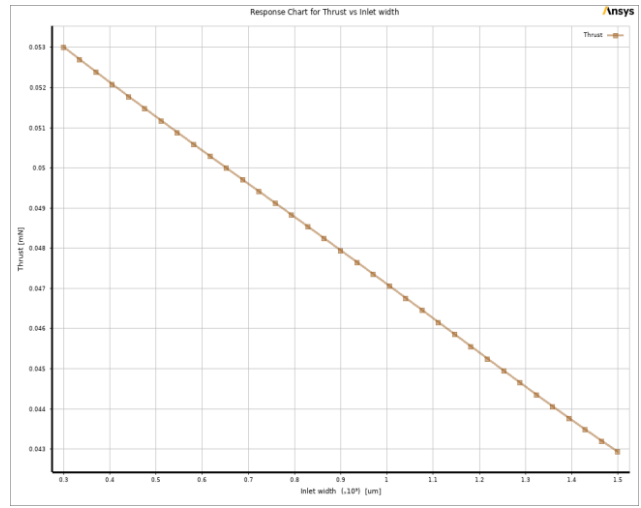

**Figure 9** The relationship between Thrust variation and the Inlet width of the C-D micro-nozzle.

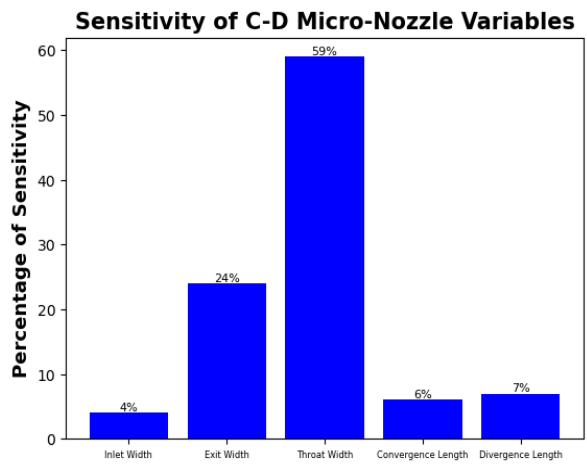

**Figure 8** The bar chart illustrates the sensitivity of various variables (Inlet width, exit width, Throat width, Convergence length, and Divergence length) concerning the C-D micro-nozzle.

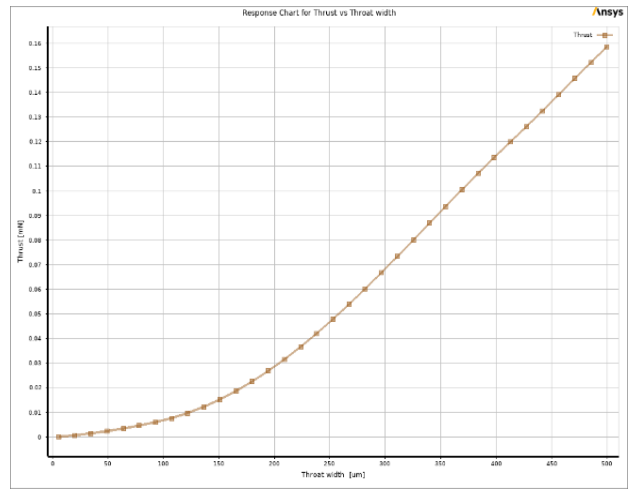

**Figure 10** The relationship between Thrust variation and the Throat width of the C-D micro-nozzle.

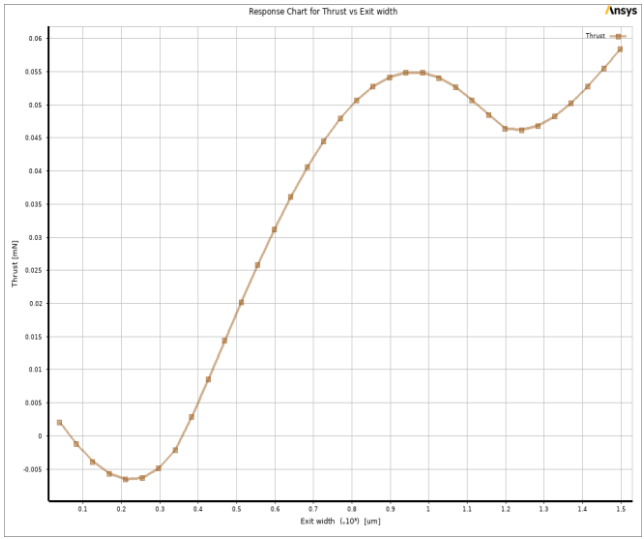

**Figure 11** The relationship between Thrust variation and the Exit width of the C-D micro-nozzle.

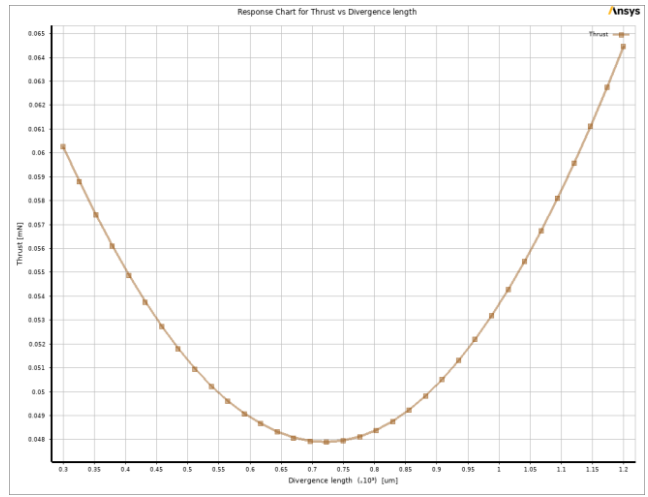

**Figure 13** The relationship between Thrust variation and the Divergence length of the C-D micro-nozzle.

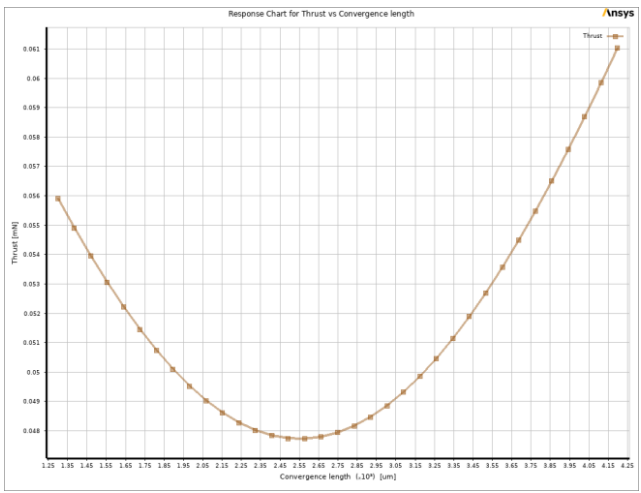

**Figure 12** The relationship between Thrust variation and the Convergence length of the C-D micro-nozzle.

In figure 14-18, to facilitate a more comprehensive analysis, the generated thrusts have been plotted as 3D surface plots against the geometric variables of the C-D micro-nozzle. These plots enable a comparative evaluation of the influence of each variable on the generated thrust in relation to another variable.

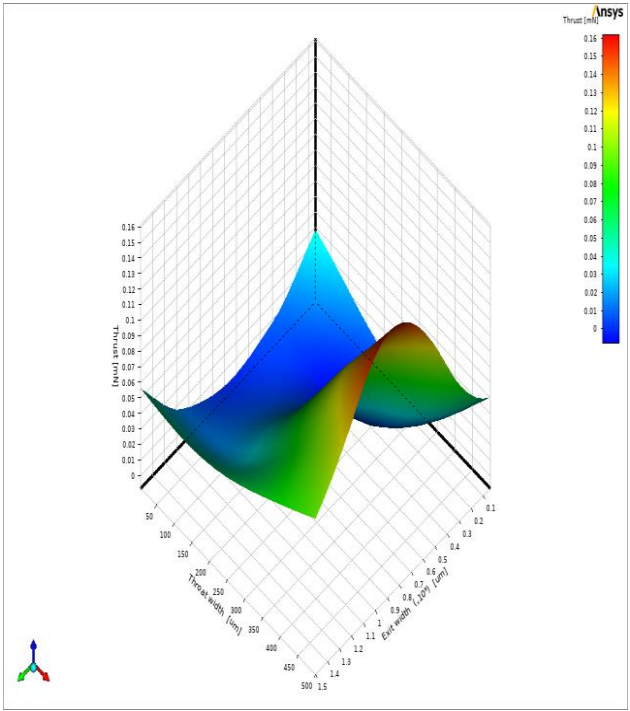

(a) Thrust variation as a function of the Throat width and the Exit width of the C-D micro-nozzle

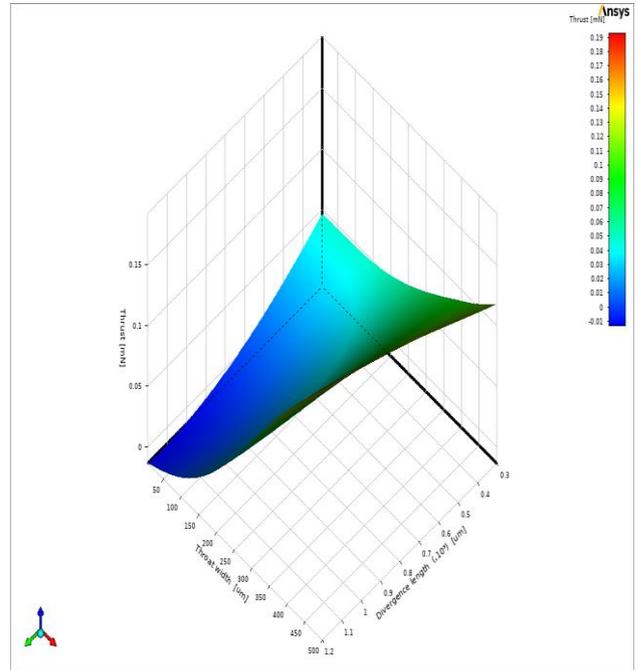

(a) Thrust variation as a function of the Throat width and the Divergence length of the C-D micro-nozzle

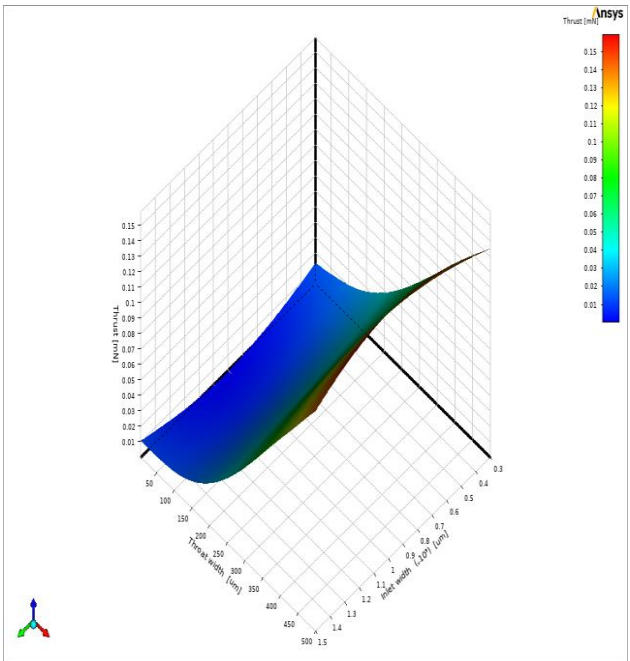

(b) Thrust variation as a function of Throat width and the Inlet width of the C-D micro-nozzle

**Figure 14** 3D surface plots of thrust.

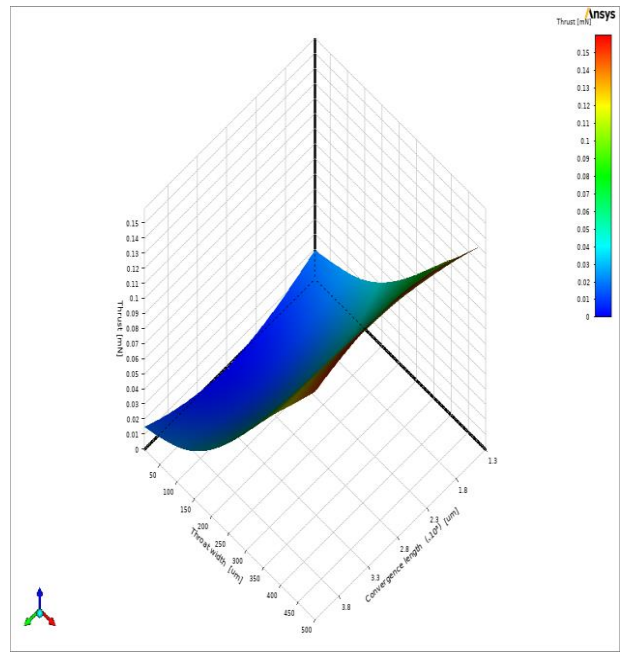

(b) Thrust variation as a function of Throat width and the Convergence length of the C-D micro-nozzle

**Figure 15** 3D surface plots of thrust.

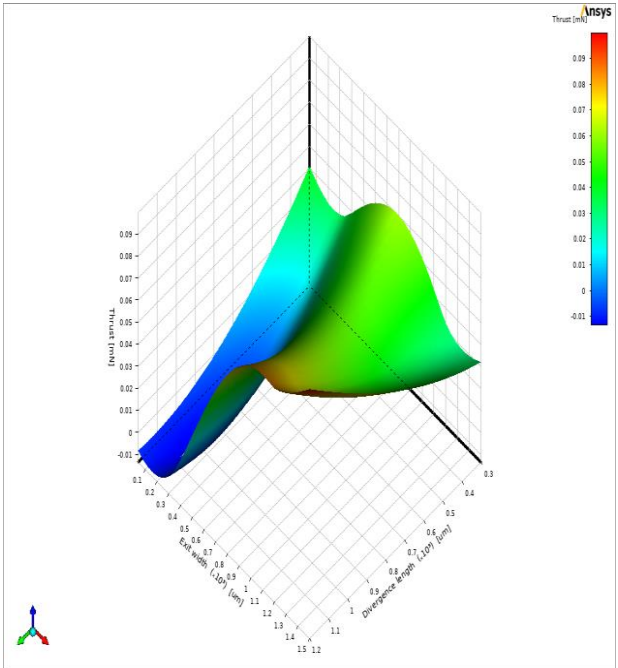

(a) Thrust variation as a function of the Exit width and the Divergence length of the C-D micro-nozzle

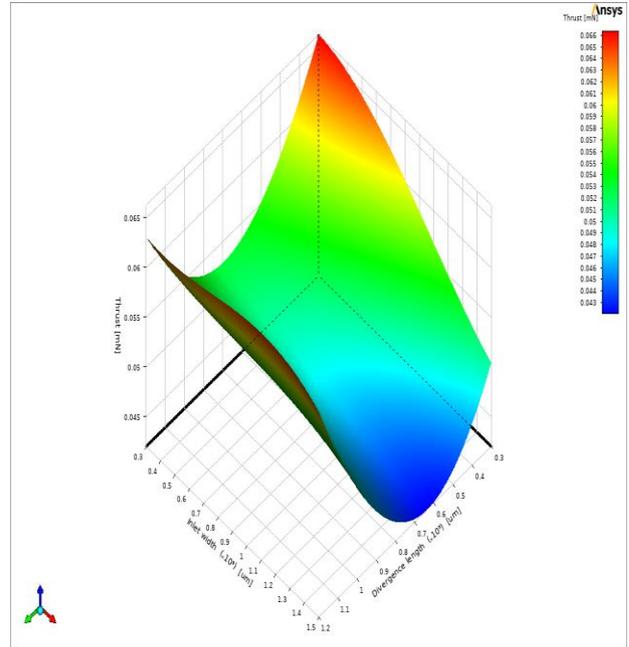

(a) Thrust variation as a function of the Inlet width and the Divergence length of the C-D micro-nozzle

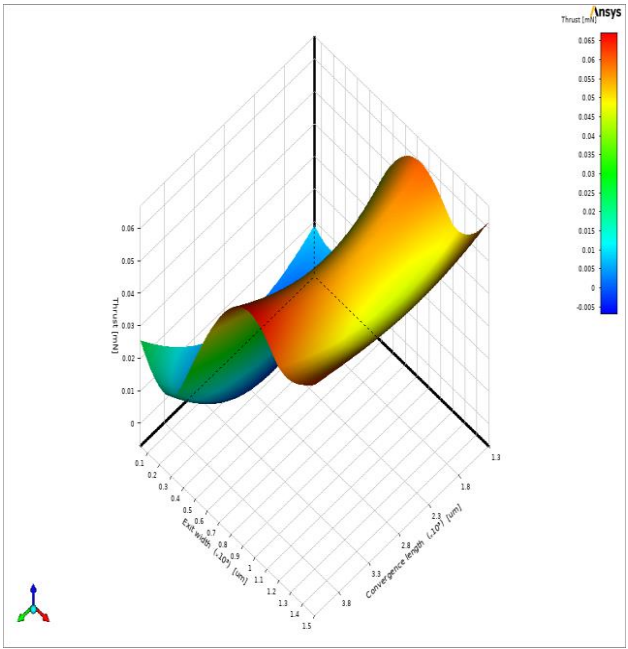

(b) Thrust variation as a function of Exit width and the Convergence length of the C-D micro-nozzle

**Figure 16** 3D surface plots of thrust.

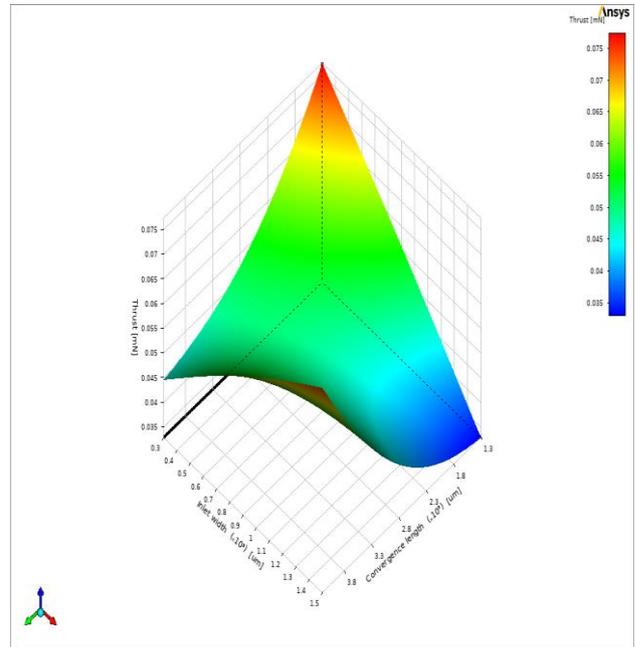

(b) Thrust variation as a function of Inlet width and the Convergence length of the C-D micro-nozzle

**Figure 17** 3D surface plots of thrust.

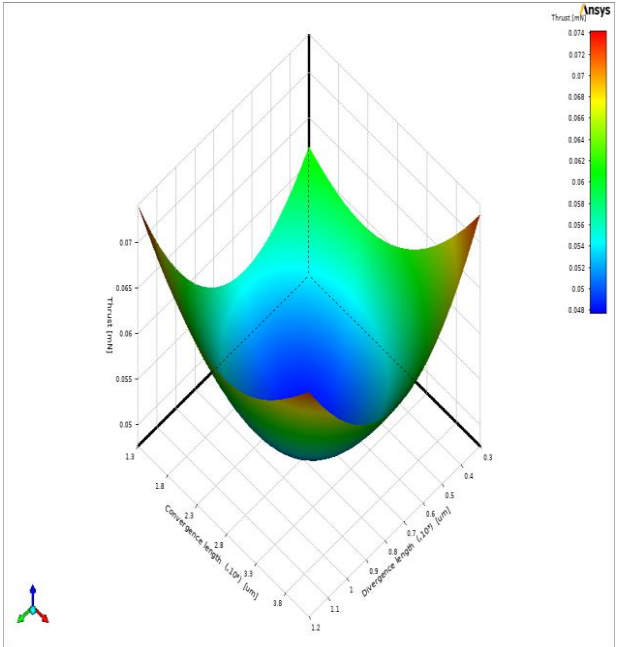

(a) Thrust variation as a function of the Convergence length and the Divergence length of the C-D micro-nozzle

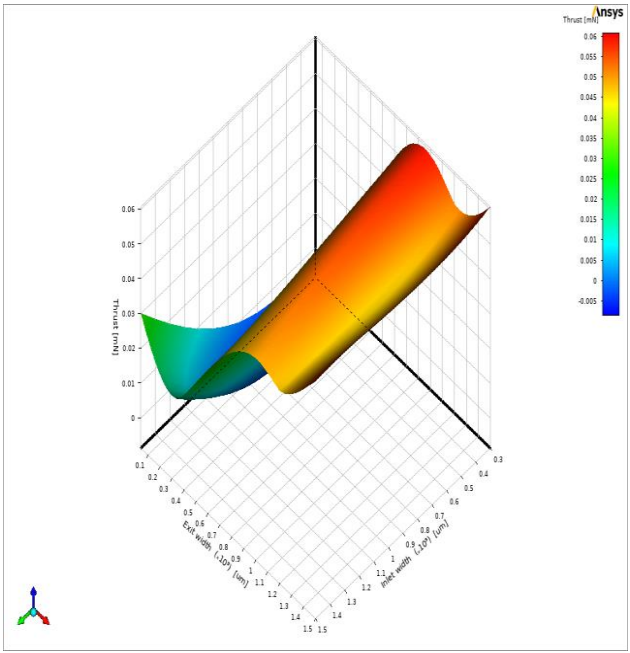

(b) Thrust variation as a function of Exit width and the Inlet width of the C-D micro-nozzle

**Figure 18** 3D surface plots of thrust.

## 3.3. Optimization

The design of experiments and analysis of response surfaces have led to the optimization of the desired geometry, aimed at enhancing the generated thrust. After careful evaluation, the best design has been selected, and the corresponding values are presented in Table 6. To visualize the optimized C-D micro-nozzle geometry, refer to Figure 19. Notably, the application of response surface optimization has resulted in a substantial increase in thrust, reaching 113.1 mN, which holds great significance for the micro thruster.

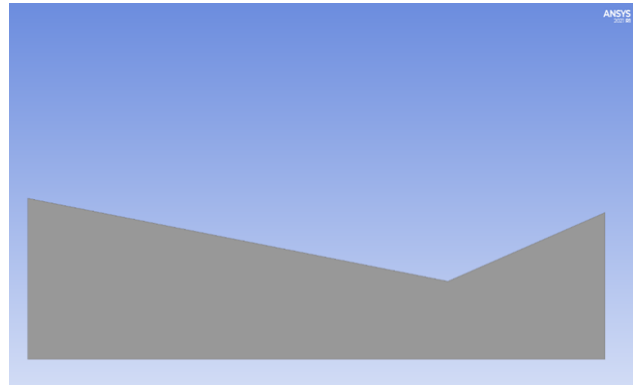

**Figure 19** Schematic of the optimized geometry of C-D micro-nozzle using Response Surface Optimization method.

## 3. Numerical Results for Curved-Throat C-D Micro-Nozzle Performance

In the simulation and optimization process, the idea of applying a curved throat for the nozzle was examined. It means that in the optimized design from the previous section using response surface optimization, the normal throat (Figure 19) was replaced by the curved throat (Figure 20). This idea of replacement, caused around 25 percent increase in the thrust to achieve 141 mN which surprisingly enhances the efficiency of C-D micro thruster without any change in the size of the C-D micro-nozzle.

It has been suggested in the literature, that the best throat radius is usually equal to the diameter of the throat [22] so according to Table 7, the simulation was conducted. Schematic of C-D micro-nozzle geometry with a curved throat using a radius of 999.22 μm is shown in Figure 20. The velocity and Mach number variation start to increase along the convergence section of the C-D micro-nozzle and reach their maximum values at the throat, and then decrees after passing the throat (Figure 21). Pressure variation along the C-D micro-nozzle is shown in Figure 22 which approves the drop in pressure in the throat area. The velocity vector contour also is shown in Figure 22.b which confirms that the optimized design could eliminate the reverse flow which downright reduces the C-D micro-nozzle performance.

**Table 6** Optimal variable values for C-D micro-nozzle performance.

| Variable | Inlet width | Exit width | Throat width | Convergence length | Divergence length | Thrust |
|---|---|---|---|---|---|---|
| Optimized design | $W_i(\mu m)$ | $w_e(\mu m)$ | $W_t(\mu m)$ | $L_{conv}(\mu m)$ | $L_d(\mu m)$ | $F$ (mN) |
|  | 1031.4 | 940.04 | 499.61 | 3207 | 1199.2 | 113.1 |

**Table 7** The values of the optimized design for C-D micro-nozzle variables, incorporating a curved throat.

| Variable | Inlet width | Exit width | Throat width | Convergence length | Divergence length | Thrust |
|---|---|---|---|---|---|---|
| Optimized design | $W_i(\mu m)$ | $w_e(\mu m)$ | $W_t(\mu m)$ | $L_{conv}(\mu m)$ | $L_d(\mu m)$ | $F$ (mN) |
|  | 1031.4 | 940.04 | 999.22 | 3207 | 1199.2 | 141 |

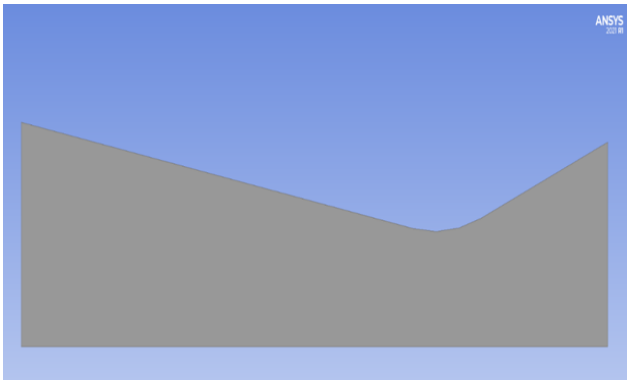

**Figure 20** Schematic of C-D micro-nozzle geometry with a curved throat with a radius of 999.22 μm.

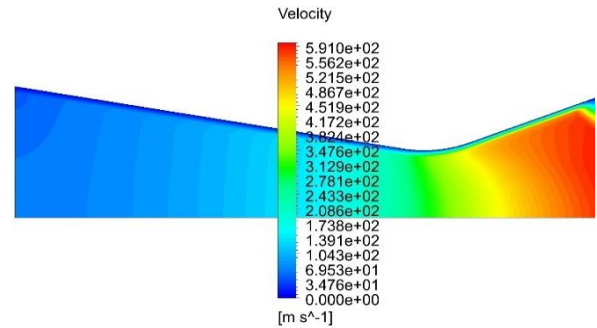

(a) velocity variation for a 2-bar pressure difference

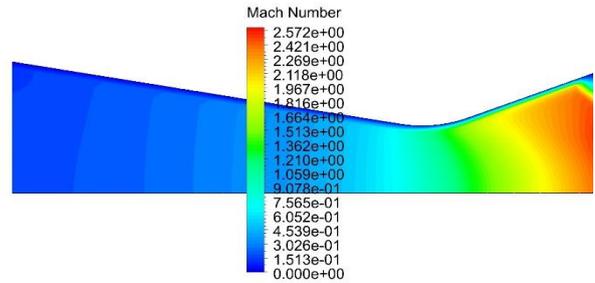

(b) Mach number variation for a 2-bar pressure difference

**Figure 21** Variation in pressure and Mach number variation for a 2-bar pressure difference under sea level conditions.

In Figure 20, the schematic of the C-D micro-nozzle geometry is presented, featuring a curved throat with a radius of 999.22 μm. The curvature in the throat section introduces a smooth and gradual transition for the fluid flow, promoting efficient acceleration and expansion. The use of a curved throat is strategically employed to enhance the performance and effectiveness of the micro-nozzle design. The chosen radius of 999.22 μm is carefully selected to achieve optimal flow characteristics and maximize thrust generation.

The velocity variation for a 2-bar pressure difference at sea level conditions is depicted in Figure 21. The color contour represents the velocity distribution across the surface of the dual-throat C-D micro-nozzle. It can be observed that the velocity increases as the fluid flows through the nozzle, reaching its maximum at the throat section. Subsequently, the velocity decreases as the fluid expands and exits the nozzle. These velocity contours provide valuable insights into the flow behavior and acceleration within the dual-throat C-D micro-nozzle.

Similarly, Figure 21 illustrates the Mach number variation for a 2-bar pressure difference at sea level conditions. The color contour displays the distribution of Mach numbers across the surface of the dual-throat C-D micro-nozzle. The Mach number is a dimensionless quantity that indicates the ratio of the fluid velocity to the local speed of sound. It offers crucial information about the compressibility and flow regime within the nozzle. From Figure 21, it can be observed that the Mach number increases as the fluid accelerates through the throat section, eventually reaching supersonic values. As the fluid expands and exits the nozzle, the Mach number gradually decreases. The Mach number contours aid in understanding the flow regime and identifying the presence of supersonic or subsonic flows within the dual-throat C-D micro-nozzle.

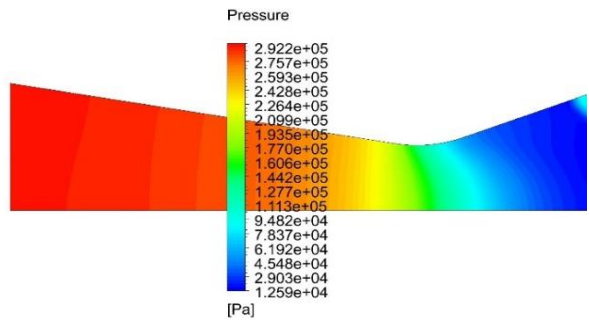

(a) Pressure variation for a 2-bar pressure difference

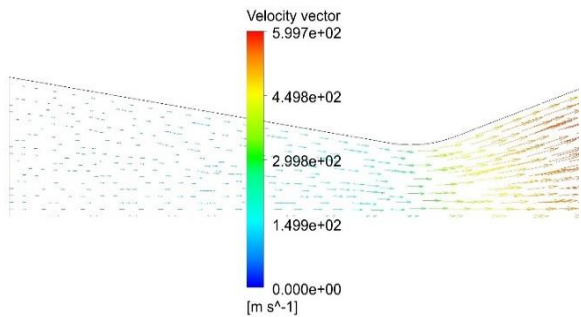

(b) Velocity vector variation for a 2-bar pressure difference

**Figure 22** Pressure Variation and Velocity vector variation for a 2-bar pressure difference under sea level conditions.

The color contour in Figure 22 represents the pressure distribution across the surface of the dual-throat C-D micro-nozzle. It is evident that the pressure gradually decreases from the inlet to the outlet of the nozzle, indicating a pressure drop as the fluid flows through the nozzle. This pressure variation is crucial in understanding the fluid dynamics and the changes occurring within the dual-throat C-D micro-nozzle.

Additionally, the velocity vectors depicted in the Figure 22 illustrate the magnitude and direction of the fluid velocity at different points along the nozzle surface. It can be observed that the fluid velocity increases as it enters the nozzle and reaches its maximum at the throat section. Subsequently, as the fluid expands and exits the nozzle, the velocity gradually decreases. The velocity vectors provide valuable insights into the flow behavior, direction, and acceleration within the dual-throat C-D micro-nozzle.

## 3. Dual-throat micro-nozzle simulation

In this section, we will investigate the dual-throat C-D micro-nozzle, which is designed to increase the thrust compared to the primary C-D micro-nozzle. The dimensions of the desired dual-throat micro-nozzle are provided in Table 8, and its geometry is illustrated in Figure 23.

Upon examining the thrust levels, we observe an increase in thrust up to 1.1 mN compared to the primary nozzle configuration. By comparing the Velocity and Mach number contours depicted in Figure 24 and Figure 25, respectively, we can observe that the presence of the second throat effectively prevents the occurrence of flow separation, which can negatively impact the performance of the C-D micro-nozzle.

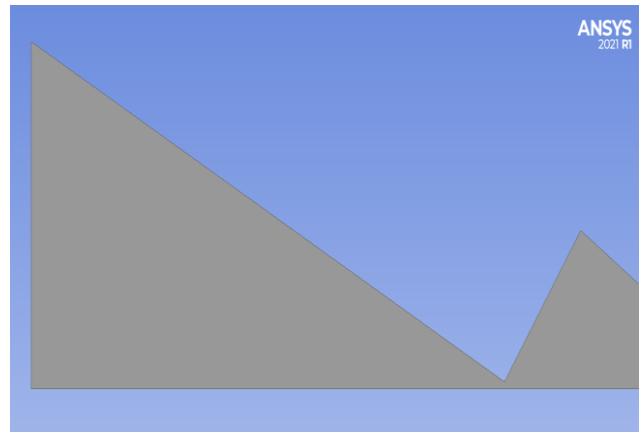

**Figure 23** The dimensions of the Dual-Throat C-D micro-nozzle geometry.

Figure 24 and 25 illustrate the geometry of the dual-throat C-D micro-nozzle. Figure 24 represents the micro-nozzle with one curved throat, while Figure 25 depicts the micro-nozzle with two curved throats, both having a radius of 60 micrometers. The micro-nozzle with one curved throat in Figure 24 generates a thrust of 2.1 mN, whereas the micro-nozzle with two curved throats in Figure 25 achieves a thrust of 2.4 mN. This further emphasizes the significance of the curved throat, as demonstrated in our findings.

**Table 8** The dimensions of the dual-throat micro-nozzle.

| Variable | Inlet width | Exit width | Throat width | Semi convergent | Semi divergent | Second-Semi convergent |
|---|---|---|---|---|---|---|
| Optimized design | $W_i\,(\mu m)$ | $w_e\,(\mu m)$ | $W_t\,(\mu m)$ | $\alpha(°)$ | $\beta(°)$ | $\alpha_2(°)$ |
| | 1500 | 450 | 30 | 28 | 54 | 36 |

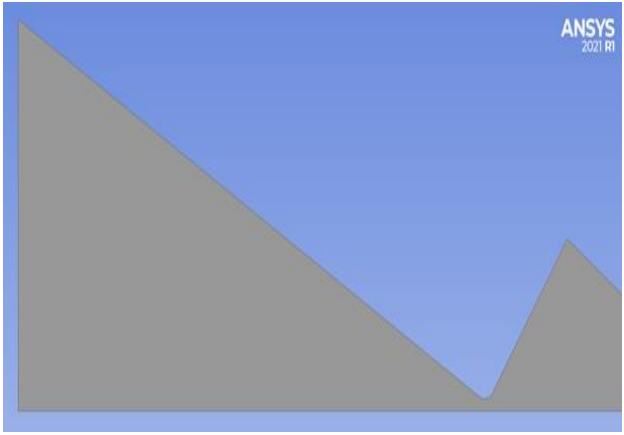

**Figure 24** The geometry of the dual-throat C-D micro-nozzle includes a curved primary throat with a radius of 60 micrometers.

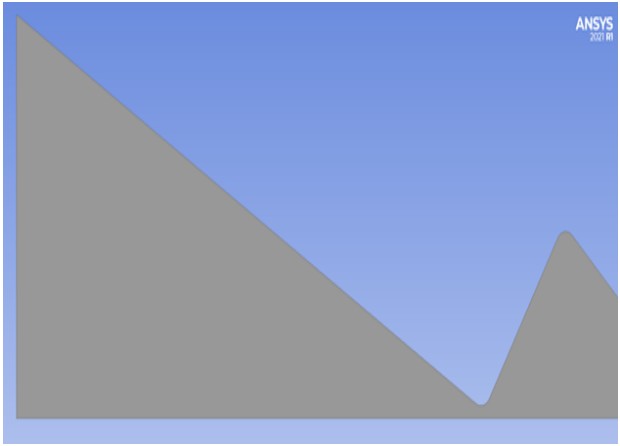

**Figure 25** The geometry of the dual-throat C-D micro-nozzle consists of two curved throats, each with a radius of 60 micrometers.

## 4. Response surface optimization for dual-throat micro-nozzle

Considering the observed increase in thrust level when using a dual-throat C-D micro-nozzle, the optimized geometry shown in Figure 26 was further optimized using the response surface optimization method. This optimization focused on the two parameters: divergence length ($L_d$) and secondary convergence length ($L_{conv_1}$), as described in Section 3. The specified ranges for these two parameters can be found in Table 9, while the response point resulting from this optimization is provided in Table 10.

**Table 9** Variable ranges for Divergence length and Secondary Convergence length of the desired dual-throat micro-nozzle.

| Parameters | Variable |
|---|---|
| Divergence length $L_d\,(\mu m)$ | 150 to 599.6 |
| Second convergence length $L_{conv_1}\,(\mu m)$ | 150 to 599.6 |

**Table 10** Parameter values for Divergence length and Secondary convergence length at the response point.

| Name | Divergence length $L_d\,(\mu m)$ | Second convergence length $L_{conv_1}\,(\mu m)$ | Thrust $F\,(\mu m)$ |
|---|---|---|---|
| Response point | 374.8 | 374.8 | 234.1 |

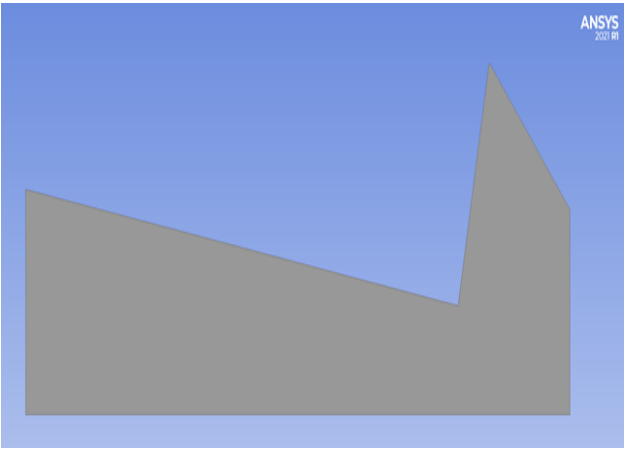

**Figure 26** Schematic of dual-throat micro-nozzle geometry optimized by Response Surface Optimization method.

After completing the optimization and validation of the candidate points, the optimal values for the two parameters, namely divergence length and secondary convergence length, are provided in Table 12. Figure 27 presents a local sensitivity analysis of these two parameters for the dual-throat C-D micro-nozzle. It demonstrates that determining the most important parameter between the two is challenging. Figure 28 and Figure 29 display the thrust diagrams corresponding to the secondary convergence length and the divergence length, respectively. Additionally, Figure 30 showcases a three-dimensional diagram illustrating the thrust variations with respect to both the second convergence length and the divergence length.

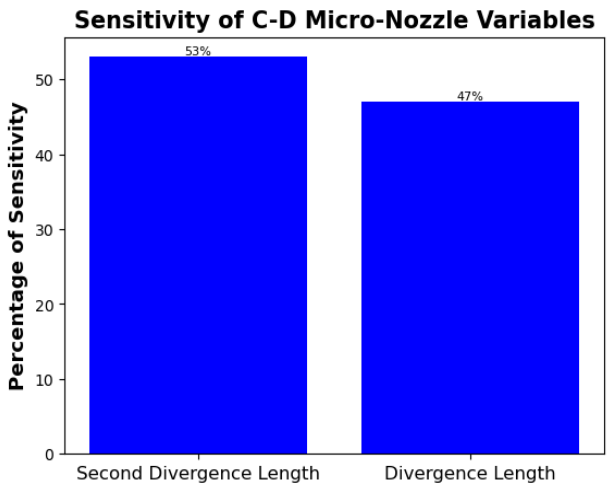

**Figure 27** Sensitivity analysis of Divergence length and Secondary convergence length for the dual-throat C-D micro-nozzle.

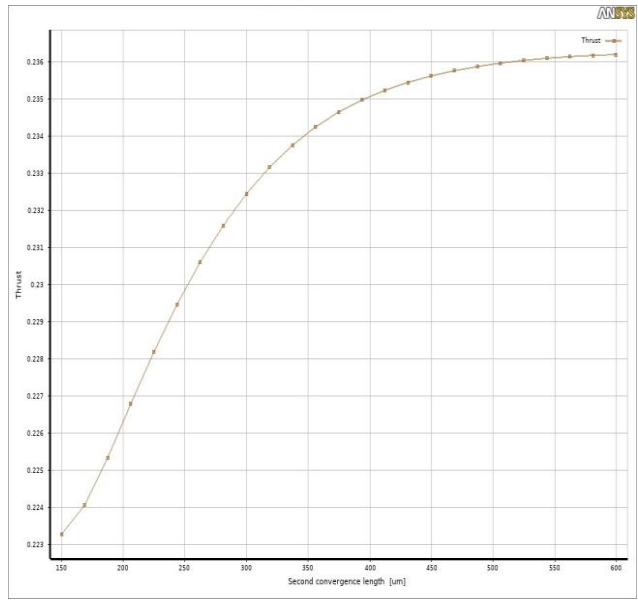

**Figure 28** Thrust variations with Second convergence length of the dual-throat C-D micro-nozzle.

;

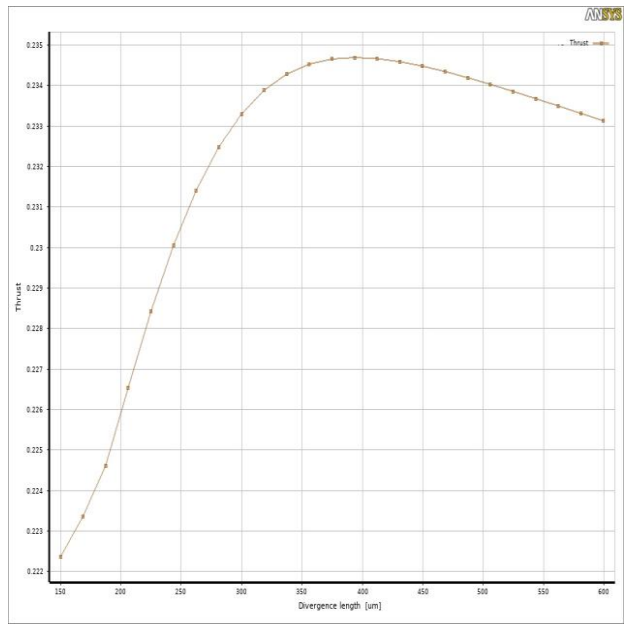

**Figure 29** Thrust variations with Divergence length of the dual-throat C-D micro-nozzle.

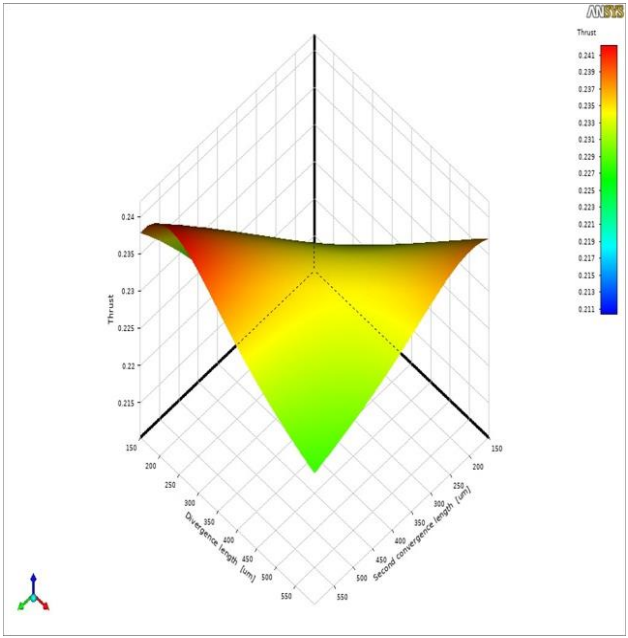

**Figure 30** 3D variation of thrust with Divergence length and Second convergence length for dual-throat C-D micro-nozzle.

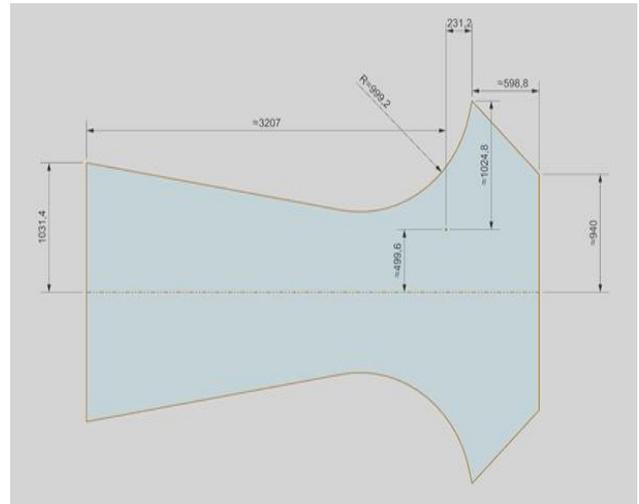

**Figure 31** Schematic of the optimized geometry of the dual-throat C-D micro-nozzle (Dimensions in micrometers).

Finally, to achieve the optimal geometry depicted in Figure 31, a curved throat design was employed. In Figure 31, the throat exhibits an initial curvature with a radius of 999.22 micrometers, resulting in a thrust of 261 mN, which significantly surpasses the performance of the geometry in Figure 26. It is worth noting that the radius of curvature should be selected within a range that does not compromise the overall geometry. While there is some flexibility in adjusting the values, empirical evidence suggests that larger curvature radii tend to increase thrust levels. Although the second throat curvature was not utilized in the final design due to its lesser impact on thrust, it remains a viable option. In the geometry illustrated in Figure 31, the initial curvature radius was set equal to the throat diameter of the dual-throat C-D micro-nozzle, based on recent experiences.

Thus, the chosen geometry for the final design is represented by Figure 31, owing to its superior thrust performance compared to the initial design. To provide a comprehensive understanding of the final design, velocity contours, Mach number distribution, static pressure distribution, and turbulence intensity are visualized in Figure 32, Figure 33, Figure 34, and Figure 35, respectively.

Please note that the visual representations in these figures enhance the comprehension of the final design and its flow characteristics.

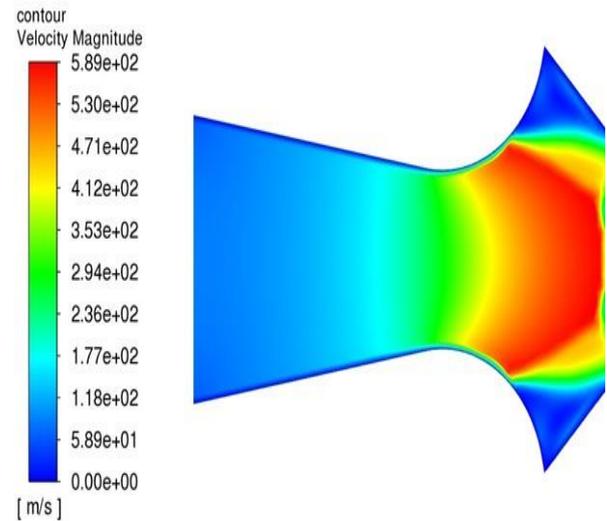

**Figure 32** Velocity Magnitude contour of the dual-throat C-D micro-nozzle.

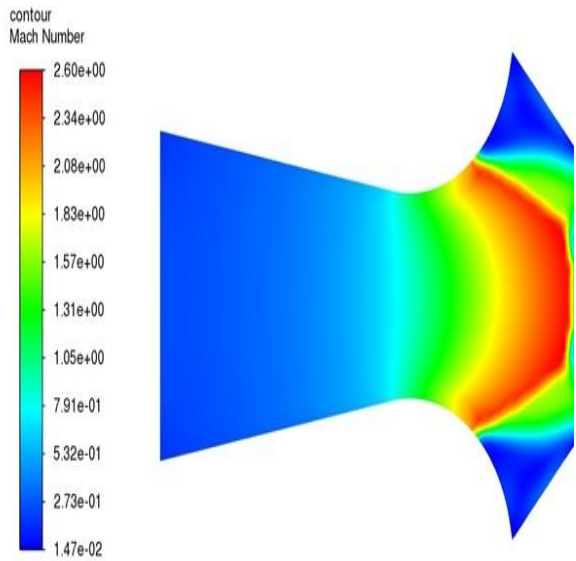

**Figure 33**  Mach Number contour of the dual-throat C-D micro-nozzle.

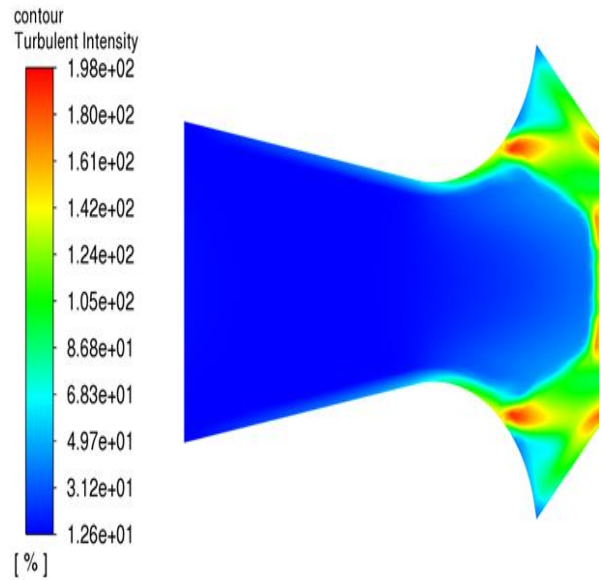

**Figure 35**  Turbulent Intensity contour of the dual-throat C-D micro-nozzle.

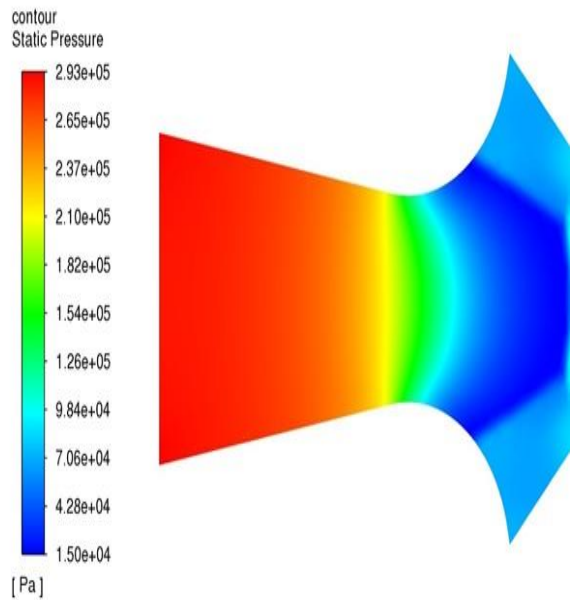

**Figure 34**  Static Pressure contour of the dual-throat C-D micro-nozzle.

## 4. Conclusion

The cold gas micro thruster is attracting significant attention due to its unique properties, including simplicity, non-toxicity, absence of combustion, and minimal leakage compared to other propulsion systems. However, optimizing the efficiency and thrust generated by micro thrusters is crucial and advantageous due to the size and mass constraints of microsatellites. In this study, a computational fluid dynamics (CFD) model of the micro-nozzle, a crucial component of the micro thruster, was developed using ANSYS FLUENT software. The simulation process was validated using experimental data from the literature, and the impact of each geometric variable on thrust generation was investigated through extensive simulations. A thrust of 0.68 mN at 2-bar feed pressure was achieved under ambient conditions. The simulations revealed the significant influence of throat width and exit width on thrust generation, outweighing the effects of other variables such as inlet width, convergence, and divergence lengths. Response surface optimization was applied to obtain the optimal geometry for the C-D micro-nozzle, resulting in a noteworthy thrust of 113.1 mN, considering the size of the thruster. Furthermore, a curved throat design was implemented, leading to a 25 percent enhancement in thrust, reaching 141 mN. Inspired by studies highlighting the potential of dual-throat micro-nozzles to increase thrust, a dual-throat configuration was simulated. Response surface optimization was performed for two important parameters, divergence length, and the second convergence length, resulting in an astonishing thrust increase up to 261 mN. This approach not only enhances thrust but also enables the utilization of less complex micro propulsion systems, leading to cost-effectiveness.

In conclusion, this work demonstrates innovation in several aspects: the utilization of a cold gas micro thruster, the development of a computational model, the investigation of geometric variables, the application of response surface optimization, and the exploration of innovative design modifications. These contributions enhance the efficiency, thrust generation, and cost-effectiveness of micro propulsion systems, thereby advancing the field of microsatellite missions.

In light of the foregoing analysis, several key recommendations emerge to optimize thrust force. The investigation into the effects of fuel type on thrust force strongly advocates for the diversification of fuel sources. It is recommended to conduct research on burner presence, taking into account power levels and spatial positioning, in order to optimize thrust force. Additionally, the proposed enhancement of the micro-nozzle's performance involves the incorporation of curvature in the second throat of the final geometry, coupled with the application of an optimization technique such as the response surface method. To achieve optimal results, geometric parameters should be tailored, considering environmental and micro-thruster geometric prerequisites, with adjustments made as needed. In the prospective realization of this design, adherence to functional requirements, material specifications, mission objectives, and design constraints — including measurements, dimensional considerations, thrust vector control, among others — is imperative to ensure the reliability and comprehensive fulfillment of all requisite criteria.

# Appendix

**Thrust,**
$$F = \dot{m}V_e + (P_e - P_a)A_e \tag{1}$$

**Specific impulse,**
$$I_{sp} = \frac{F}{\dot{m}} = C_f \times C^* \tag{2}$$

**Specific impulse at sea level,**
$$I_{sp} = V_e + (\frac{(p_e - p_a) \times A_e}{\dot{m}}) \tag{3}$$

**Thrust coefficient,**
$$C_f = \sqrt{\gamma} \times (\frac{2}{\gamma+1})^{(\frac{\gamma+1}{2(\gamma-1)})} \times \sqrt{\frac{2\gamma}{\gamma-1}(1-(\frac{p_e}{p_c})^{(\frac{\gamma-1}{\gamma})})} \tag{4}$$

**Characteristic velocity,**
$$C^* = \frac{\sqrt{RT_C}}{\sqrt{\gamma} \times (\frac{2}{\gamma+1})^{(\frac{\gamma+1}{2(\gamma-1)})}} \tag{5}$$

**Expansion ratio,**
$$\in = \frac{(\frac{2}{\gamma+1})^{(\frac{1}{\gamma-1})} \times (\frac{p_c}{p_e})^{(\frac{1}{\gamma})}}{\sqrt{\frac{\gamma+1}{\gamma-1}(1-(\frac{p_e}{p_c})^{(\frac{\gamma-1}{\gamma})})}} \tag{6}$$

**Jet/exit velocity,**
$$V_e = \sqrt{\frac{2 \times \gamma \times R_o \times T_c}{(\gamma-1)M}(1-(\frac{p_e}{p_c})^{(\frac{\gamma-1}{1})})} \tag{7}$$

# References

[1] C. I. Underwood, G. Richardson, and J. Savignol, "In-orbit results from the SNAP-1 nanosatellite and its future potential," *Philosophical Transactions of the Royal Society of London. Series A: Mathematical, Physical and Engineering Sciences,* vol. 361, no. 1802, pp. 199-203, 2003.

[2] L. Meena, M. Niranjan, G. Kumar, and M. Zunaid, "Numerical study of convergent-divergent nozzle at different throat diameters and divergence angles," *Materials Today: Proceedings,* vol. 46, pp. 10676-10680, 2021.

[3] R. Ranjan, K. Karthikeyan, F. Riaz, and S. Chou, "Cold gas propulsion microthruster for feed gas utilization in micro satellites," *Applied Energy,* vol. 220, pp. 921-933, 2018.

[4] B. Little and M. Jugroot, "Investigation of an electrospray within a cold gas nozzle for a dual-mode thruster," in *53rd AIAA/SAE/ASEE Joint Propulsion Conference*, 2017, p. 5038.

[5] T. Wekerle, J. B. Pessoa, L. E. V. L. d. Costa, and L. G. Trabasso, "Status and trends of smallsats and their launch vehicles—An up-to-date review," *Journal of Aerospace Technology and Management,* vol. 9, pp. 269-286, 2017.

[6] J. N. Easley, J. Young, M. Priddy, and H. Doude, "Additive Manufacturing of Propellant Tank and Structural Supports of CubeSat Cold Gas Propulsion System," in *AIAA Propulsion and Energy 2019 Forum*, 2019, p. 4309.

[7] M. Fatehi, M. Nosratollahi, A. Adami, and S. H. Taherzadeh, "Designing space cold gas propulsion system using three methods: Genetic algorithms, simulated annealing and particle swarm," *International Journal of Computer Applications,* vol. 118, no. 22, 2015.

[8] S. Igarashi, K. Yamamoto, A. B. Fukuchi, H. Ikeda, and K. Hatai, "Development Status of the 0.5 N Class Low-Cost Thruster for Small Satellite," in *2018 Joint Propulsion Conference*, 2018, p. 4753.

[9] T. D. Holman and M. Osborn, "Numerical optimization of micro-nozzle geometries for low reynolds number resistojets," in *51st AIAA/SAE/ASEE Joint Propulsion Conference*, 2015, p. 3923.

[10] R. Y. Chiang, A. Wu, C. J. Dennehy, and A. A. Wolf, "Application of micro-thrusters for space observatory precision attitude control," 2021.

[11] Z. You, *Space microsystems and micro/nano satellites*. Butterworth-Heinemann, 2017.

[12] E. Buchen and D. DePasquale, "2014 nano/microsatellite market assessment," *SpaceWorks Enterprises,* vol. 12, 2014.

[13] C. Williams, B. Doncaster, and J. Shulman, "Nano/microsatellite market forecast," *SpaceWorks Enterprises Inc, Atlanta, Georgia,* 2018.

[14] E. Kulu, "Nanosatellite Launch Forecasts-Track Record and Latest Prediction," 2022.

[15] A. Davani and P. D. Ronney, "Optimal design of nozzles for microsatellite propulsion," in *AIAA Scitech 2021 Forum*, 2021, p. 0992.

[16] K. Karthikeyan, S. Chou, L. Khoong, Y. Tan, C. Lu, and W. Yang, "Low temperature co-fired ceramic vaporizing liquid microthruster for microspacecraft applications," *Applied Energy,* vol. 97, pp. 577-583, 2012.

[17] T. Croteau and A. D. Greig, "Micro-Nozzle Design for an Electrothermal Plasma Thruster," in *AIAA Propulsion and Energy 2019 Forum*, 2019, p. 4243.

[18] J. Mueller *et al.*, "Towards micropropulsion systems on-a-chip: initial results of component feasibility studies," in *2000 IEEE Aerospace Conference. Proceedings (Cat. No. 00TH8484)*, 2000, vol. 4: IEEE, pp. 149-168.

[19] M. A. Silva, D. C. Guerrieri, A. Cervone, and E. Gill, "A review of MEMS micropropulsion technologies for CubeSats and PocketQubes," *Acta Astronautica,* vol. 143, pp. 234-243, 2018.

[20] M. de Athayde Costa e Silva, "MEMS Micropropulsion: Design, Modeling and Control of Vaporizing Liquid Microthrusters," 2018.

[21] A. Fluent, "ANSYS Fluent Theory Guide, ANSYS," *Inc., Release,* vol. 15, 2013.

[22] M. C. Louwerse, "Cold gas micro propulsion," *University of Twente,* p. 176, 2009.